Citation:

# IEEE-GDL CCD Smart Buildings Introduction

V.M. Larios, J.G. Robledo, L. Gómez, and R. Rincon

*IEEE Guadalajara Phisical Infraestructure Working Group for Smart Cities*

**Abstract**—As part of the activities of the IEEE-GDL CCD working group of physical infrastructure, this whitepaper is intented to be an initial guide to understand the layers, taxonomy of services and best practices for the development of smart buildings. Open standards are claimed in order to increase interoperability between layers and services. Moreover, two buildings in Guadalajara city, one new and another to renew, are described as a proof of concept under development and being part of the strategy to develop the smart city infrastructure based in a master plan. A discussion will be addressed in order to identify the areas of innovation and opportunities for the smart buildings as the contribution of this document.

——————————  ◆  ——————————

## 1 INTRODUCTION

THIS document intends to introduce the basic concepts of the technological infrastructure required for a smart building. The infrastructure is referenced with the international best practices to apply in the urban renovation of Guadalajara City and the goal to develop its smart city approach. Rather than just a set of best practices and standards referenced in this document, a smart building is a key component of the smart city and potential testbed for the IT infrastructure and city solutions to scale in public areas.

There are many documents already developed that explain the concepts regarding a smart building, and moreover, the concept is evolving. Another value of this whitepaper stands to stablish a definition and identify the different layers of the technology in a building referecing the components and associated standards as well as relevant building references. We also propose two projects under development at Guadalajara City to follow, the first one is a new building and the second one is a renewal building where the certifications and practices proposed in this whitepaper are meant for implementation. Both buildings are part of the catalyst projects in the Guadalajara city masterplan strategy for the Digial Creative City, developed by Professors Carlo Ratti and Dennis Frenchman from the MIT Senseable Lab as project consultants [1].

## 2 CONTEXT

### 2.1 Smart Building Definition

A smart building is a construction with an appropiate desing and technological support to maximize its functionalities and confort for their occupants with the compromise to reduce their operational costs, and extend the life of the physical structure [2]. The smart buildings adapted in a local environment look to optimize four basic correlated elements [3]:
1) Physical structure
2) Systems
3) Services
4) Management

The figure 1 shows the relationship of the building elements.

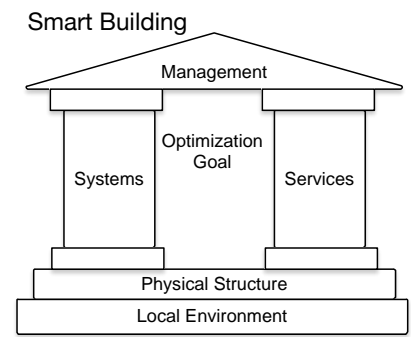

Fig. 1. The four correlated elements to optimize in a smart building.

A building can offer many services loking to increase the confort, productivity and security of their inhabitants. The services are related to the management of systems integrating sensors and actuators on the fisical infrastructure adapted to the local environment where the building

————————————————

- *V.M. Larios is a Full Professor at the Universidad de Guadalajara CUCEA, Information Systems Department and the Director of the Smart Cities Innovation Center at the same institution. E-mail: vmlarios@cucea.udg.mx*
- *J.G. Robledo is Researcher at the Smart Cities Innovation Center at CUCEA Universiddad de Guadalajara. E-mail: jrobledo@cucea.udg.mx.*
- *L. Gomez is a Researcher at the Smart Cities Innovation Center at CUCEA Universidad de Guadalajara, E-mail: lgomez@cucea.udg.mx*
- *R.Rincon is a Graduated Student from ITESO University, Guadalajara, E-mail: ti672055@iteso.mx*

is placed [4].

In a smart city, the buildings are the first cells and the sum of many smart buidings can be seen as an important part of a sustainable urban environment. Hence, the optimal goals of the smart buidings are dealing with a better use of natural resources and also are part of a complex system from a systemic approach[5].

Moreover, from the technological point of view, we can identify the different layers of a smart building in more detail, to understand the correlation of the systems, services, and management operations. For each layer, is important to understand the implied actors, stakeholders and best practices to implement that we will discuss in the following sections.

### 2.2 Layers of a Smart Building

For a first approach we can identify three main layers in a smart building related to the physical structure, connectivity and the software, as shown in figure 2.

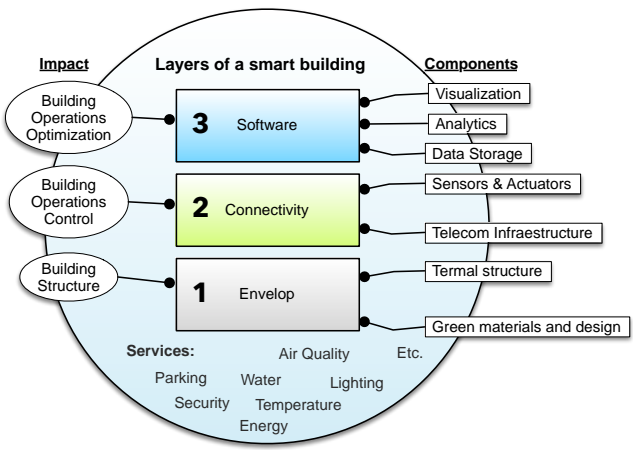

Fig. 2. The layers of a smart building.

The envelop of the building is related to all the materials and a green desing of architects and civil engineers looking to make a more efficient the use of resources as water, lighting, security, temeperature, air quality, parking, etc. This first layer is the body of the building and is related with the materials and design conceived by the architects for the productive activities of the inhabitants. The connectivity is the nervous system the building and comprises the data network connecting all the sensors and actuators to react with the environment. In the connectivity layer, open standards as the TCP/IP protocol for a better software integration is recommended for interoperability of systems [6].
Finally the software layer is the brain of the building connected by the nervous system and using its body to interact with the people inside and outside.

Furthermore, the software can be divided in sublayers as shown in figure 3.

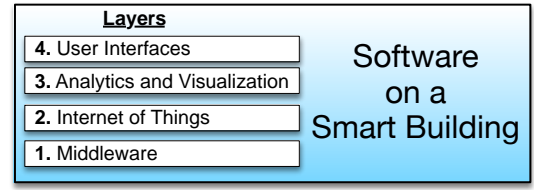

Fig. 3. Software layers inside a building.

The Middlewere is related to the low level programing as firmware on embedded systems used to read sensors and control actuators. The Internet of Things relates to the specific software and services created to coordinate and manage the different interconnected objects, identified as part of the building. All the sensors provide information to the analytics platform in order to find the patterns, models and reach the optimal point of operation of each subsystem in the building. Finally, the interfaces are the specific settings of the building inhabitants based on their activities.

The past classification of layers is important to understand the dinamics of a building and their different elements in order to identify the best practices related to certifications and standards available when a building is planned to be intelligent. We will have more information related to standards in section 3, afther looking in more detail the different subsystems in the taxonomy of a smart building.

### 2.3 Taxonomy of basic services in a smart building

From the layers of a smart building there are many integrated services that can be seen as subsystems. The set of services are managed to provide the best conditions for the activities of the building ocupants. Hence, the figure 4 shows the taxonomy of basic services.

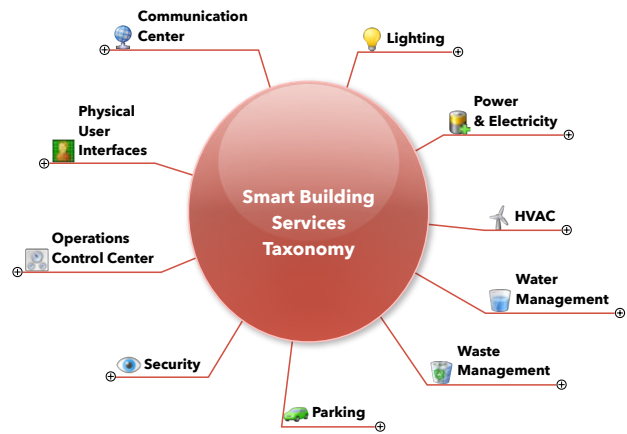

Fig. 4. Smart Building Services Taxonomy.

All the services have sensors and actuators in order to manage their operation. Lighting service is related for the confort of people depending their activities and has sensors to save energy when lights are not needed. Some buildings have on their envoltory a system to filter the natural light using automated curtains in order to save energy. The power and electricity system should have on-site renewable energy sources as solar photovoltaic panels to supply a percentage of the power used in the building services [7], [8].

HVAC is related to the humidity, ventilation and air conditioning system, designed for the confort of users and with an effective interaction with the environment. The water management service look to promote savings, to threat and reuse wastewater as possible for flushing, landscaping and aircooling systems. The waste management is related to the strategy to collect and separate efficiently. As an example, the building can use organic waste as fertilizer in the green areas as well as implement a recycling program for plastics and metals [9].

Parking service should promote car sharing, electric vehicles and a place for bicycles. The security is important in all the building, managing automated locks, biometric devices as well as video surveillance systems. The operations control center is the place where all the analytics of systems are executed and the decisions to support the opperatinos are taken. Visual interfaces integrate a dashboard of the building services showing the status and must support human operators to better manage the building resources. The physical user interfaces connects the people with the building to set up their optimal parameters for confort and productivity improvement on daily activities. Finally, the communications center is the core of the network connecting sensors and actuators in the building as well as the operations control center. These are the basic set of services that are found in almost every smart building today [10].

## 3 REVIEW OF STANDARDS FOR SMART BUILDINGS

Around the world we see many success cases and best practices that have been used for the construction of intelligent buildings [11],[12]. We will review them in our layers proposed structure. For the envenlope layer, there are two main certifications related with the envoiement that are the LEED [13] and BREEAM [14]. LEED is from the U.S. Green Building Council (USGBC) in Washington. BREEAM is BRE Environmental Assestment Method stablished in the UK by the Building Research.

Some organizations as the International Telecomunications Union (ITU) created a technical reference document for sustainability explaining deeper this standards as part of the toolkit on environmental sustaniabilty for the ICT sector [15]. Both LEED and BREEAM propose a rating system for the design, construction and operation of buildings. The estandards look for high performance of systems and green sustainability. They evaluate building projects based in categories as water effiency, energy and atmosphere, materials and resources, indoor environmental quality and innovation in design. For LEED a building

TABLE 1
ENVELOP LAYER STANDARDS

| Process or Technology Regulation | Description |
|---|---|
| **ISO TC 205** Building environment design | Generally includes the building indoor environmental quality, caring the energetic efficiency, air quality, sonority and visual factors. |
| **IEC 62305-4** Protection against lightning for electrical and electronic systems within structures for smart buildings. | This standard establishes all the rules to protect users, electronic and electrical equipment against lightning discharges in thunderstorms. It becomes important to have the considerations and safety issues established in this standard. |
| **ISO 16484-1:2010** Project specification and implementation for building automation and control systems. | It is a guide in the design of the building project that especially impacts on control systems and automation peer. Standard as should be the design documents with the terms of technical reference, the engineerings with detailed hardware functions, facilities, documentation and training. It becomes relevant to CCD to narrow this standard tenders. |

*These standards are complementary to LEED and BREEAM certifications.*

TABLE 2
CONNECTIVITY LAYER STANDARDS

| Process or Technology Regulation | Description |
|---|---|
| **ISO 16484-2:2004** Building automation and control systems hardware | Provides the norm for identification of all associated hardware elements to the control and automation for the building, establishing practices for interfaces, automation stations of specific drivers, cabling and devices interconectiion, engineerings and necessary tools. This standard is complementary to physical devices cabling for data networks in ISO 16484-5. |
| **ISO 16484-6:2009** Building automation and control systems data communication conformance testing | Defines a standard methodology to ensure that all automation system and building control protocols implemented fulfill what they promise to control. In itself, this standard seeks to review each service and really have a control system established as an initiator process, connection and implemented functions for efficient control operations. |
| **ISO 16484-5:2012** Building automation and control systems data communication protocol | Defines data communication services and protocols for computer equipment used to monitor and control heating, air conditioning, ventilation systems and other building systems. Helps to define in an object-oriented paradigm the representation of information associated with each control equipment to facilitate the development of control digital applications in buildings. |

*These standards establish the best practices for buildings automation for the services of the building. It is important to look for open standards in order to allow interoperability between layers.*

TABLE 3
SOFTWARE LAYER STANDARDS

| Process or Technology Regulation | Description |
|---|---|
| **ISO 16484-3:2005** Building automation and control systems functions | Sets a standard of documentation for the functions and engineering of control systems and intelligent building automation. Includes main requirements and software application functions to control and operate automated processes of the building. |

*Other standards from the software industry could apply in a more general point of view.*

can be certified with a regular label, silver, gold and platinum depending on the ratings for each category. Usually LEED gold certification is a good compromise for the envelope of a smart building in terms of green sustainability. In the case of BREEAM, similar categories are evaluated on a scale of pass, good, very good, excellent and outstanding, where the excellent level is also a good compromise for the smart buildigns. Related to the topic of energy, other certifications as Green Star could be envisioned.

In addition to the certifications, there are also some standards to review for the envelope layer as shown in table 1. The standards refers to good practices to prepare the envelope for the connectivity layer where table 2 introduces the data transport in the hardware for atomation and control systems and provide services in the building. Aditionally, it sould be considered for the software layer the following ISO 16484-3:2005 for the software related to the building automation and control functions as shown in table 3.

## 4 PROPOSAL FOR SMART BUILDINGS IN GUADALAJARA CITY

We propose to follow the development of two building as experience, a new building for working. The new building will be known as "The Digial Creative Complex" hosting a a combined diverse activities and facilities to reatail, small and medium sized businesses for the creative industries, and shared digital creative services and cafes. The building will be developed with hybrid architecture between historic and contemporary interventions and will be a model to be replicated in the future.

The renewal project is an integrated facility that will be known as "The Ingenium Campus" designed to support education. It takes place at an historic Basilio Vadillo school with an art nouveau stlyle becoming the vector of knowledge of the city. At the Ingenium institute, it is envisioned as a joint venture of educational institutions with a board of schools in the metropolitan area of Guadalajara and industry members. The principal role will be to serve as a link to the industry and support joint programs and projects. The potential areas of focus for the

TABLE 4
SMART BUILDINGS PROJECTS IN GUADALAJARA CITY

| Features | First creative new building | Building renewal. The Ingenium Campus |
|---|---|---|
| Surface | 25,000 m$^2$ | 5,282 m$^2$ |
| Estimated building architectural structure | Three towers of 5 floors each interconnected with underground parking on two levels with 400 car coverage. | A complex of buildings with classrooms, laboratories, libary, adminsitration, auditorium and sports facilities. |
| Users | 2500 | 1000 |
| Users profile | Office, research laboratories associated to the digital creative sector, meeting rooms/auditorium, shopping mall. | Office, research laboratories associated to the digital creative sector. |

*Both buildings are in process of development and construction according to the Gudalajara CCD Master Plan.*

Ingenium are envisioned: educational multi-media using the Internet, smart cities with a living lab and media and the arts to encourage creative applications and artistitc expressing throught digital means. Table 4 presents some information about the buildings currently in process of design and construction to be acomplished by the end of 2015. The buildings will be certified as LEED Gold and support presented on this withepaper as proof of concepts.

## 5 DISCUSSION AND PERSPECTIVES

We presented a model of layers and services for smart buildigns and grouped standars related to such layers. A basic set of services was described and two proposed projects in progress as part of the experience of the Gudalaraja smart city are told. For next steps, this document shoud be feed with the final output of the buildings and the learned lessons. Other works in progress as the living lab to experiment and test products for the smart city may enrich this proposal. An area of opportunity identified on this project is first, the use of smart buildings as a testbed for the emerging projects from the living labs under development. In particular, the interfaces to interact with the building and to measure the user experience.

**ACKNOWLEDGMENT**

The authors wish to thank the Gudalajara CCD organization for their support, in particular Octavio Parga as the president of the organization for sharing its vision and supporting the working groups initiative. Also, we thank the IEEE Guadalajara section volunteers, CANIETI Occi-

LARIOS ET AL.: IEEE-GDL CCD SMART BUILDINGS INTRODUCTION 5

dente and PROMEXICO for their support. A special thank is for ITESO University and their different instances as: Coordinación de Proyectos de Aplicación Profesional, CPAP (Carlos F. Ruiz Sahagún), Oficina de Sistemas de Información, OSI (Eduardo M. Chihenseck Leal), Oficina de Servicios Generales (Luis Fernando Orozco Cabrera), PAP professors Araceli García Gómez and Victoria E. Espinoza Cabrera. Finally, we appreciate the support of the Universidad de Guadalajara and its PhD in IT program providing students as well as the advices of Intel, IBM and HP on the working groups.

**V.M. Larios** has received his PhD in 2001 and a DEA in 1996 (French version of a MS program) in Computer Science at the Technological University of Compiègne, France and a BA in Electronics Engineering in 1996 at the ITESO University in Guadalajara, Mexico. He works at the University of Guadalajara (UDG) as Professor Researcher since 2001 and as consultant directed the Guadalajara Ciduad Creativa Digital Science and Technolgy program during 2013. His research insteres are related to distributed systems, visual simultaitons and smart cities. He is a Senior IEEE member and current chair of the Computer Chapter at the IEEE Guadalajara Section at Region 9. His role in the IEEE-CCD Smart Cities initiative is to lead the working groups.

**J.G. Robledo** ose Robledo's Communications and Electronics Engineer graduated from the University of Guadalajara. Subsequently obtained his Masters degree in Business Administration at the University Center for Economic and Administrative Sciences UDG and currently is a PhD candidate in Information Technologies (DTI ) on the topic called " Crowdsourcing for Smart Cities : Improving Data Quality and Trustworthiness of Data Sources".He has contributed to propel the Exploration Center Solutions IBM Smart Cities - UDG developing prototypes of mobile applications and simulations for the implementation of a Smart Campus in CUCEA UDG . During 2013 he has been participating in the Guadalajara-CCD project as internship at the Science & Technology Office. The areas of interest of M. Jose Guadalupe Robledo is currently focused on issues Crowdsourcing Simulation of Multi- Agent Systems in the context of Smart Cities , Visual Analytics and BigData Georeferenced and Predictive Analytics . José Robledo is a member of the Computer Society at IEEE and an active volunteer in the Internet of Things working group from the IEEE-CCD Smart City Initiative.

**L. Gomez** has received his PhD in IT recently in January 2014 and a Master in IT in 2006, and a BA in Computer Science in 2000 all at the Universidad de Guadalajara. During 2013 he has been participating in the Guadalajara-CCD project as internship at the Science & Technology Office. His research interests are related to systems optimization, metaheuristitcs, and smart cities. He is an IEEE member and active volunteer in the Internet of Things working group from the IEEE-CCD Smart City Initiative.

**R. Rincon** is an undergraduated student at ITESO University in the bachelor in IT program on its fourt and last year. Member of IEEE since june 2013 is an active volunteer at the physical infrastructure working group from the IEEE-CCD Smart City initiative.